\documentclass[pdflatex,sn-mathphys-num]{sn-jnl}
\usepackage{graphicx}%
\usepackage{multirow}%
\usepackage{amsmath,amssymb,amsfonts}%
\usepackage{amsthm}%
\usepackage{mathrsfs}%
\usepackage[title]{appendix}%
\usepackage{xcolor}%
\usepackage{textcomp}%
\usepackage{manyfoot}%
\usepackage{booktabs}%
\usepackage{algorithm}%
\usepackage{algorithmicx}%
\usepackage{algpseudocode}%
\usepackage{listings}%

\theoremstyle{thmstyleone}%
\theoremstyle{thmstyletwo}%

\theoremstyle{thmstylethree}%

\DeclareMathOperator\sign{sgn}

\begin{document}

\title[Strong potential in a box for applications to femtoscopy]{Strong potential in a box for applications to femtoscopy}


\author[1,2]{\fnm{Gleb} \sur{Romanenko}}\email{gleb.romanenko2@unibo.it}

\author[1,2]{\fnm{Francesca} \sur{Bellini}}\email{f.bellini@unibo.it}

\affil[1]{\orgdiv{Departmemt of Physics and Astronomy "Augusto Righi"}, \orgname{University of Bologna}, \orgaddress{Via Irnerio 46, \city{Bologna}, \postcode{40126}, \country{Italy}}}

\affil[2]{\orgname{Istituto Nazionale di Fisica Nucleare}, \orgaddress{\city{Bologna}, \postcode{40126}, \country{Italy}}}

\abstract{Understanding the short-range nucleon–nucleon interaction is essential for the interpretation of correlation femtoscopy measurements in high-energy hadronic and nuclear collisions. We present an analytical treatment of the strong interaction in two-nucleon systems by modeling it with a square-well potential and solving the Schrödinger equation in the presence of the Coulomb interaction. The resulting pair wave function is regular at small relative distances and allows for the inclusion of multiple partial waves. 
We apply this framework to proton–proton femtoscopy and compute theoretical correlation functions for realistic source sizes. We demonstrate that the commonly used Lednicky–Lyuboshits asymptotic approximation overestimates the correlation signal for small sources. Comparisons with numerical calculations using the CATS framework and the Argonne \textit{v}18 potential show good agreement within current experimental uncertainties. The proposed analytical approach provides a practical and flexible tool for femtoscopic analyses of nucleon and baryon pairs.}




\maketitle

\section{Introduction}\label{sec1}

One of the long-standing open problems in nuclear physics is the precise formulation of the strong interaction potential between two nucleons. Even for the simplest case of a nucleon–nucleon (N--N) pair, a fully controlled derivation of the interaction from Quantum Chromodynamics (QCD) remains elusive. First-principles calculations are currently feasible only within the restricted framework of effective field theory (EFT), and therefore much of our quantitative understanding still relies on phenomenological potential models. 
In these models, nucleon–nucleon scattering data and information extracted from the properties of light nuclei are used to build empirical parameterizations of the strong potential. This strategy has led to successful and widely used formulations, such as Nijm93, Reid93~\cite{PhysRevC.49.2950} and Argonne $v18$~\cite{PhysRevC.51.38}, among others.

Beyond its intrinsic theoretical relevance, knowledge of the strong interaction at short distances is also crucial for understanding the formation and structure of nuclear bound states. For example, the precise measurements of (anti)deuteron production at the Large Hadron Collider (LHC)~\cite{ALICE:2019dgz, ALICE:2020foi, ALICE:2020hjy} have stimulated new developments in coalescence models \cite{Butler:1963pp, Kapusta:1980zz}, where the N--N interaction enters as a key input to describe the formation of light nuclei in high-energy collisions~\cite{Kachelriess:2019taq, Mahlein:2023fmx, Mahlein:2024pur, Leung:2025jwe}. 

A natural conceptual bridge between the structure of nuclear bound states and two-hadron interactions is provided by correlation femtoscopy \cite{Blum:2019suo, Mrowczynski:1992gc, Bellini:2020cbj, Scheibl:1998tk}.
Historically related to Hanbury–Brown--Twiss interferometry \cite{Brown:1956zza}, it is considered to provide another experimental avenue for accessing short-distance physics. 
Femtoscopy is widely used by experiments \cite{Wiedemann:1999qn} at the Super Proton Synchrotron (SPS)~\cite{NA61SHINE:2023qzr}, the Relativistic Heavy-Ion Collider (RHIC)~\cite{PHENIX:2015jaj, Gos:2007cj} and the Large Hadron Collider (LHC)~\cite{ALICE:2011dyt, ALICE:2015hvw, ALICE:2020mkb} to study the space-time properties of the particle-emitting source. 

The method of femtoscopy relies on measuring momentum correlations between particle pairs arising from quantum statistics and final-state interactions (FSI)~\cite{Lisa:2005dd}. While the Coulomb interaction and the statistical symmetrization or antisymmetrization are well understood, the description of the strong FSI remains uncertain due to its short range, which is difficult to probe directly.
\\The shape of a two-particle correlation function is determined by the emission source profile as well as by the interaction between the particles in a pair, embedded in the pair wave function (WF).
In the Koonin–Pratt formula \cite{Koonin:1977fh, Pratt:1990zq}

\begin{equation}
    C(k, R_{\mathrm{eff}}) = \int \mathrm{d}^3 \Vec{r} \; S(\Vec{r}, R_{\mathrm{eff}}) \; {\left| \Psi (\Vec{r}, \Vec{k}) \right|}^2,
    \label{eq_CF_th}
\end{equation}

\noindent the source function $S(\Vec{r}, R_{\mathrm{eff}})$ defines the probability for a pair to be emitted with a given relative pair distance $\Vec{r}$ from a source with an effective size $R_{\mathrm{eff}}$, $\Psi (\Vec{r}, \Vec{k})$ is the pair WF and $\Vec{k} = \frac{1}{2} (\Vec{p}_2 - \Vec{p}_1)$ is the half of the pair relative momenta in their Center-of-Mass (CM).
\\Consequently, femtoscopy can be used in two complementary ways: (i) if the interaction is known, femtoscopy provides access to the spatial and temporal characteristics of the emitting source \cite{Lisa:2005dd}, (ii) if the source is known or constrained independently, the measured correlations allow one to test or constrain the interaction potential. The latter has motivated the application of femtoscopy to systems created in high-energy collisions as they offer a unique environment to produce a broad set of particle species. For instance, this was applied at the LHC to study correlations between nucleons and strange hadrons with low relative momenta, gaining new insights on the strong interaction potential at the femtometer scale ~\cite{Fabbietti:2020bfg, ALICE:2020mfd, STAR:2025jwe}.
At the same time, because the femtoscopic signal manifests itself at low relative momentum (i.e. for $\lesssim 200$ MeV/$c$), some authors have argued whether femtoscopy is sensitive at all to the short-range strong interaction \cite{Epelbaum:2025aan}.

For nucleon systems, we are interested in finding an effective formulation of accounting for the strong potential that can be used in femtoscopic analyses with the purpose of measuring the nucleon-emitting source size, test its scaling properties across momentum and collision systems or multiplicities, and possibly, apply this experimental knowledge to study the nuclei formation through coalescence \cite{Mahlein:2023fmx, Bellini:2020cbj}.

In this work we focus on modeling the short-range strong interaction through a square-well potential. This allows us to obtain an analytic solution of the Schr\"odinger equation for a pair of identical charged fermions interacting via both Coulomb and an arbitrary short-range potential. We apply this formalism to identical-proton femtoscopy, compute the corresponding theoretical correlation functions, and compare our results with existing models in the field~\cite{Mihaylov:2018rva, Lednicky:2005tb}. We conclude by discussing the immediate applicability of our solution to femtoscopic analyses in high-energy hadronic and nuclear collisions.

\section{Solution for the WF}\label{sec2}

To solve the quantum mechanical problem for a pair of interacting particles it is convenient to approach it in the CM frame where one can utilize the the spherical symmetry of the interaction potentials (Coulomb and strong). In the CM the problem is equivalent to the one of a single particle scattering on an effective center (coinciding with the CM) resulting into the reduction of the dimensionality of the problem from six (three per particle) to just three \cite{Messiah}. The corresponding Schr\"odinger equation:

\begin{equation}
    \left[ - \frac{\hbar^2}{2 \mu} \nabla^2 + \hat{V} (\Vec{r}) \right] \psi (\Vec{r}) = E \psi (\Vec{r}) ,
    \label{eq_Schred_gen}
\end{equation}

\noindent where $\mu = \frac{m_1 m_2}{m_1 + m_2}$ is the reduced mass of a pair and $E = \frac{k^2 \hbar^2}{2 \mu}$ with $k = |\Vec{k}|$.

Because the system possesses spherical symmetry, the orbital angular momentum is conserved and the solution $\psi (\Vec{r})$ can be split into angular and radial parts \cite{Messiah, Landau:1991wop, SpringerBook} as 

\begin{equation}
    \psi (\Vec{r}) = \frac{1}{k r} \sum_{l=0}^\infty i^l (2l + 1) e^{i \sigma_l} u_l (r) P_l (\cos \theta),
    \label{eq_sol_gen}
\end{equation}

\noindent where $\sigma_l$ is the phase shift for an orbital momentum of value $l$, $\theta$ is the angle between $\Vec{k}$ and $\Vec{r}$ and $u_l (r)$ is the solution to the radial Schr\"odinger equation

\begin{equation}
     \frac{\mathrm{d}^2 u_l (r)}{\mathrm{d} r^2} + \left[ k^2 - \frac{l (l + 1)}{r^2} - \frac{2 \mu}{\hbar^2}  V(r) \right] u_l (r) = 0.
    \label{eq_Schred_rad_1}
\end{equation}

\paragraph{Coulomb potential}
Solving Eq.~(\ref{eq_Schred_rad_1}) for the case of the Coulomb potential alone (see Appendix~\ref{secA1}) one obtains a general regular form of the WF, $\psi_c$, in terms of partial waves (\ref{eq_sol_gen}) that satisfies (\ref{eq_Schred_gen}):

\begin{equation}
    \begin{split}
        & \psi_c (\eta, \; \rho) = \frac{1}{\rho} \; \sum_{l=0}^\infty (2l + 1) \; i^l e^{i \sigma_l} F_l (\eta, \rho) \; P_l (\cos \theta) \equiv \\
        & \equiv \frac{1}{2 \rho} \; \sum_{l=0}^\infty (2l + 1) \; i^{l+1} \Big( u_l^- (\eta, \rho) -  e^{2i \sigma_l} u_l^+ (\eta, \rho) \Big) \; P_l (\cos \theta),
    \end{split}
    \label{eq_Coul_sol_gen}
\end{equation}

\noindent where $\rho = kr$, $\eta = \frac{1}{k a_B}$, $a_B = \frac{\hbar^2}{\mu Z_1 Z_2 e^2}$ is the Bohr radius of a pair with the reduced mass $\mu$ and charges $Z_1$ and $Z_2$, $\sigma_l = \arg \: \Gamma(l + 1 + i \eta)$ are the Coulomb phase shifts, $F_l (\eta, \rho)$ is the \textit{regular Coulomb wave function} and $u_l^{\pm} (\eta, \rho)$ are the \textit{irregular Coulomb wave functions} (\ref{eq14}). 

\paragraph{Strong potential in the asymptotic region} 
\label{subsec21}
A short-range potential behaving as $V^{s}(r) \sim o (r^{-2})$ at sufficiently large distance $r>\tilde{d}$ can be introduced to the solution given by Eq. (\ref{eq_Coul_sol_gen}) by adding a corresponding phase shift $\delta_l$ \cite{Messiah}. Thus, in the asymptotic region the solution $\psi_{c+s}$ for the Coulomb plus a short-range $V^{s}(r)$ has the following form:

\begin{equation}
    \psi_{c+s} (\eta, \; \rho) \underset{r > \tilde{d}}{=} \frac{1}{2 \rho} \; \sum_{l=0}^\infty (2l + 1) \; i^{l+1} \Big( u_l^- (\eta, \rho) -  e^{2i \sigma_l} e^{2i \delta_l} u_l^+ (\eta, \rho) \Big) \; P_l (\cos \theta).
    \label{eq_Coul_str_asympt}
\end{equation}

Solution (\ref{eq_Coul_str_asympt}) can be also rewritten through the $F_l$ and $G_l$ functions (\ref{eq14}):

\begin{equation}
    \psi_{c+s} (\eta, \; \rho) \underset{r > \tilde{d}}{=} \frac{1}{r} \; \sum_{l=0}^\infty (2l + 1) \; i^l e^{i \sigma_l} \Big( \frac{F_l (\eta, \rho)}{k} + f_l (k) \; \big(G_l (\eta, \rho) + i \: F_l (\eta, \rho) \big) \Big) \; P_l (\cos \theta),
    \label{eq_Coul_str_asympt_Led}
\end{equation}

\noindent where $f_l (k)$ is the scattering amplitude \cite{Messiah, Landau:1991wop}:

\begin{equation}
    f_l (k) = \frac{e^{2 i \delta_l} - 1}{2 i k} = \frac{1}{k \cot \delta_l - i k}.
    \label{eq_Scat_ampl}
\end{equation}

\paragraph{Short-range potential as a square-well}\label{subsec22}

\begin{figure}[h]
    \centering
    \begin{minipage}{0.48\linewidth}
        \includegraphics[scale=0.08]{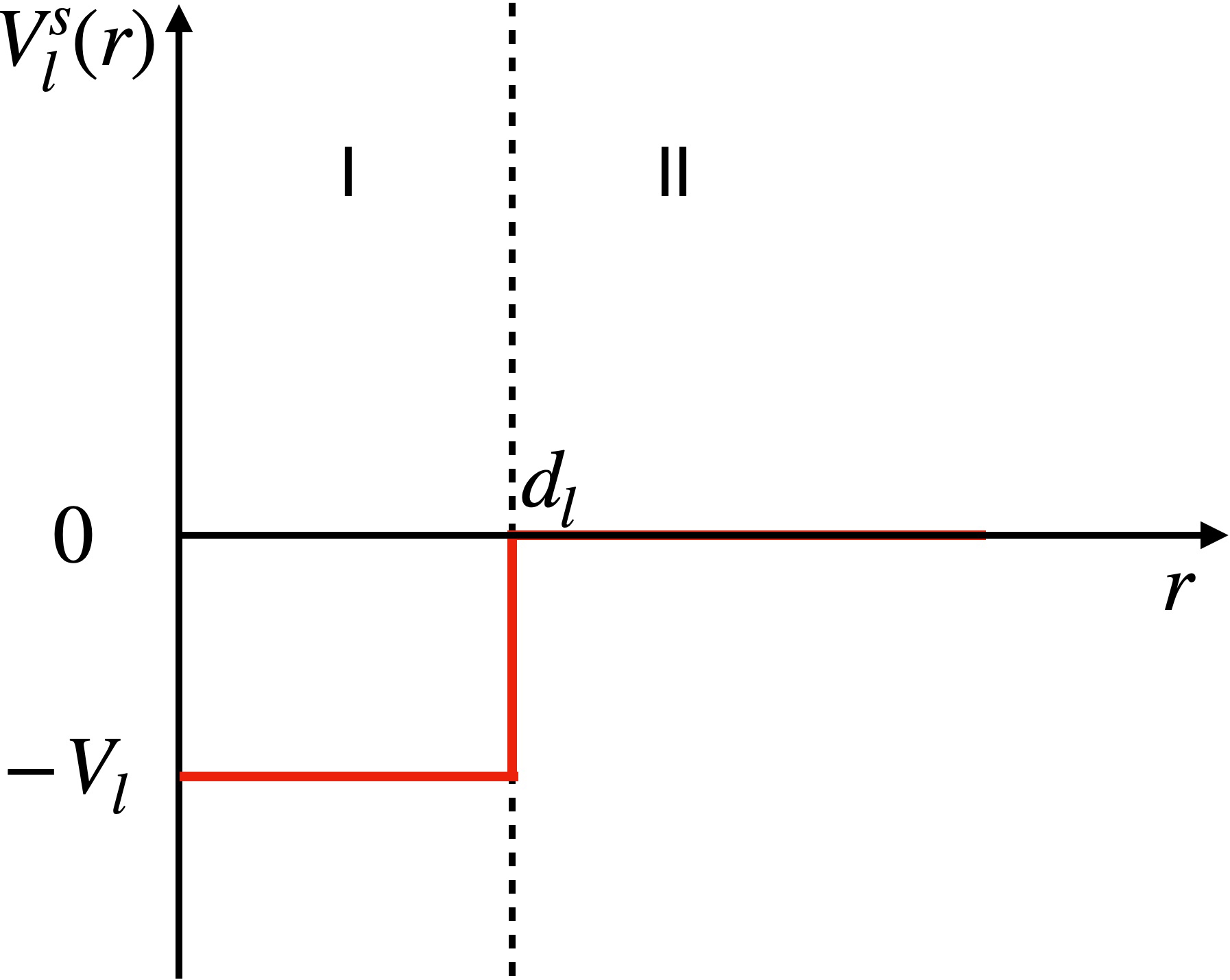}
    \end{minipage}
    \begin{minipage}{0.48\linewidth}
        $V_l^s (r) =
            \begin{cases}
                V_l, \;\;\;\;\; r < d_l \;\;\;\;\;\; (I) \\
                0, \;\;\;\;\;\; r \geq d_l \;\;\;\;\;\; (II) \\
            \end{cases}$ 
    \end{minipage}
    \caption{Square-well potential.}
    \label{fig1}
\end{figure}

In this work the strong potential is effectively considered in the shape of a square-well (SW) with different widths ($d_l$) and depths ($V_l$) for different values of $l$. This accounts for the dependence on $l$ that is observed in phase-shift experimental data \cite{PhysRevC.48.792}.
An illustration of the SW potential is reported in Fig. \ref{fig1}.

The solutions for Coulomb+SW are found separately for the $r < d_l$ region (I) of the SW influence and the asymptotic region $r \geq d_l$ (II). To find the total solution, these are then matched at $r=d_l$. The solution in region (II) is already known as in Eq. (\ref{eq_Coul_str_asympt} -\ref{eq_Coul_str_asympt_Led}), thus only the solution for region (I) needs to be found. 
To take into account the square-well potential in this region, we add a corresponding  term $\frac{2 \mu}{\hbar^2} \; V_l$, which is then substituted by the energy term $k^2$ since the potential is constant. The radial Schr\"odinger's equation in region (I) is given as

\begin{equation}
    \frac{\mathrm{d}^2 u_l^{(I)}}{\mathrm{d} r^2} +\left[ \tilde{k}_l^2 - \frac{l(l+1)}{r^2} - \frac{2}{a_B r} \right] u_l^{(I)} = 0
    \label{eq_Schred_reg_1}
\end{equation}

\noindent with $\tilde{k}_l = \sqrt{k^2 - \frac{2 \mu}{\hbar^2} \; V_l}$.

Noticing that the Eq. (\ref{eq_Schred_reg_1}) has the same form as the equation (\ref{eq_Schred_rad_1}) for Coulomb, one immediately writes the WF similarly to (\ref{eq_Coul_sol_gen}):

\begin{equation}
    \psi_{c+s}^{(I)} (\tilde{\eta}_l, \; \tilde{\rho}_l) = \frac{1}{\tilde{\rho}_l} \; \sum_{l=0}^\infty (2l + 1) \; i^l e^{i \tilde{\sigma}_l} F_l (\tilde{\eta}_l, \tilde{\rho}_l) \; P_l (\cos \: \theta),
    \label{eq_Coul_str_sol_reg1}
\end{equation}
\noindent where $\tilde{\eta}_l = \frac{1}{\tilde{k}_l a_B}$, $\tilde{\rho}_l = \tilde{k}_l r$ and $\tilde{\sigma}_l = \arg \: \Gamma(l + 1 + i \tilde{\eta}_l)$.

The total solution is then obtained by matching the solutions in Eqs. (\ref{eq_Coul_str_sol_reg1}) and (\ref{eq_Coul_str_asympt} -\ref{eq_Coul_str_asympt_Led}) for regions (I) and (II), respectively, as detailed in Appendix \ref{secA2}. One obtains

\begin{equation}
    \begin{split}
        & \psi_{c+s} (k, \; r) = \frac{1}{r} \; \sum_{l=0}^\infty (2l + 1) \; i^l e^{i \sigma_l} \;  u_l (k, \; r) \; P_l (\cos \: \theta) \\
        & u_l (k, \; r) =
    \begin{cases}
        \frac{F_l (\tilde{\eta}_l, \; \tilde{k}_l r)}{F_l (\tilde{\eta}_l, \; \tilde{k}_l d)} \; \Big( \frac{F_l (\eta, k d)}{k} + f_l (k) \; \big(G_l (\eta, k d) + i \: F_l (\eta, k d) \big) \Big), \;\;\;\;\;\; r < d_l \\
        \Big( \frac{F_l (\eta, \rho)}{k} + f_l (k) \; \big(G_l (\eta, \rho) + i \: F_l (\eta, \rho) \big) \Big), \;\;\;\;\;\;\;\;\;\;\;\;\;\;\;\;\;\;\;\;\;\;\;\;\;\;\; r \geq d_l.
    \end{cases}
    \end{split}
    \label{eq_Coul_str_sol_all_waves}
\end{equation}

As the centrifugal term $\frac{l(l+1)}{r^2}$ in the Schr\"odinger Eq. (\ref{eq_Schred_rad_1}) is zero only for $l=0$ and grows rapidly with $l$, it is reasonable to account for the short-range potential only for a finite number of partial waves $l = \{0 .. n\}$ \footnote{If $l > n$ then $\Tilde{k} \equiv k \implies \Tilde{\rho} \equiv \rho$ and $\Tilde{\eta} \equiv \eta$  as well as $f_l (k) = 0$.}. Thus, the solution (\ref{eq_Coul_str_sol_all_waves}) can be rewritten as

\begin{equation}
    \begin{split}
        & \psi_{c+s} (k, \; r) = \sqrt{A_c (\eta)} \; e^{i \sigma_0} e^{i \Vec{k} \Vec{r}} \; _1F_1 \Big(- i \eta, \; 1, \; i (kr - \Vec{k} \Vec{r}) \Big) + \\
        & + \sum_{l=0}^n (2l + 1) \; i^l e^{i \sigma_l} a_l (k, \; r) P_l (\cos  \theta), \\
        & a_l (k, \; r) =
    \begin{cases}
        \frac{F_l (\tilde{\eta}_l, \; \tilde{k}_l r)}{F_l (\tilde{\eta}_l, \; \tilde{k}_l d)} \left[ \frac{F_l (\eta, k d)}{k r} + f_l (k) \; \frac{G_l (\eta, k d) + i \: F_l (\eta, k d)}{r} \right] - \frac{F_l (\eta, \rho)}{k r}, \;\;\;\; r < d_l \\
        f_l (k) \; \frac{G_l (\eta, \rho) + i \: F_l (\eta, \rho)}{r}, \;\;\;\;\;\;\;\;\;\;\;\;\;\;\;\;\;\;\;\;\;\;\;\;\;\;\;\;\;\;\;\;\;\;\;\;\;\;\;\;\;\;\;\;\;\;\;\;\;\;\;\;\;\;\;\; r \geq d_l.
    \end{cases}
    \end{split}
    \label{eq_Coul_str_sol_finite_waves}
\end{equation}

In the latter passage we used the partial-wave expansion of the Coulomb wave function \cite{Messiah}:

\begin{equation}
    \frac{1}{k r} \; \sum_{l=0}^\infty (2l + 1) \; i^l \; e^{i \sigma_l} F_l (\eta, \rho) \; P_l (\cos \: \theta) = \sqrt{A_c (\eta)} \; e^{i \sigma_0} e^{i \Vec{k} \Vec{r}} \; _1F_1 \Big(- i \eta, \; 1, \; i (kr - \Vec{k} \Vec{r}) \Big) ,
    \label{eq_Coul_reg_sum}
\end{equation}

\noindent where $A_c (\eta) = \frac{2 \pi \eta}{e^{2 \pi \eta} - 1}$ is the Gamow factor (Coulomb penetration factor).
\newline

Being a solution for the general case of both Coulomb and strong interaction, Eq. (\ref{eq_Coul_str_sol_all_waves}) covers also the case of proton-neutron and neutron-neutron interactions (strong alone). If one (or both) charges is zero (resulting in $a_B \to \infty\implies \eta = 0$) the Coulomb WFs are reduced to the Bessel functions \cite{Messiah, Abramowitz_Stegun:1972} and the corresponding WF is obtained.

\section{Application to femtoscopy and discussion}

The goal of this work is to explore the feasibility of using the obtained analytical effective solution Eq. (\ref{eq_Coul_str_sol_finite_waves}) in describing $N-N$ correlations. To check this we use the solution to construct the proton pair WF needed to calculate a theoretical femtoscopic correlation function as given by Eq. (\ref{eq_CF_th}). Averaging over the possible spin and spin-orbital states one writes a general expression for a given number of pair total orbital mumentum $L_\mathrm{max}$ states,

\begin{equation}
    C_{pp}(k, \; R_\mathrm{eff}) = \sum_{S=0}^1 \omega_S \sum_{L=0}^{L_\mathrm{max}} \sum_{J=|L - S|}^{L+S} \omega_{LJ} \int \mathrm{d}^3 \Vec{r} S(\Vec{r}, R_\mathrm{eff}) {\left| \frac{\psi (- \Vec{k}, \Vec{r}) + {(-1)}^S \psi (\Vec{k}, \Vec{r})}{\sqrt{2}} \right|}^2 ,
    \label{eq_CF_th_weightened_l01}
\end{equation}

\noindent where

\begin{equation}
    \begin{split}
        & \omega_S = \frac{2 S + 1}{(2 s_1 + 1)(2 s_2 + 1)} , \\
        & \omega_{LJ} = \frac{2 J + 1}{(2 S + 1)(2 L + 1)}
    \end{split}
    \label{weights}
\end{equation}

\noindent are the weights associated to pair states with given pair spin $S$, pair orbital momentum $L$ and total pair angular momentum $J$ values.

In this work we will consider the strong potential to be relevant only in s- and p- (pair) states (i.e. $L=0$ and $L=1$ respectively). The choice of the square-well and effective range parameters for the strong interaction is discussed in Appendix \ref{secA3}. 

Starting with the s-wave state only\footnote{Here we mean that the strong potential is taken into account only for $L=0$ and neglected for higher partial waves whereas the contributions from the Coulomb are included for all $L$ (see Eq. (\ref{eq_Coul_reg_sum}))} (i.e. $L_\mathrm{max}=0$ in Eq. \ref{eq_CF_th_weightened_l01}) we utilize the general solution obtained before (\ref{eq_Coul_str_sol_finite_waves}) to write the WF used to construct the femtoscopic CF (\ref{eq_CF_th_weightened_l01}):

\begin{equation}
    \begin{split}
        & \psi_{c+s}^{l=0} (k, \; r) = \sqrt{A_c (\eta)} \; e^{i \sigma_0} e^{i \Vec{k} \Vec{r}} \; _1F_1 \Big(- i \eta, \; 1, \; i (kr - \Vec{k} \Vec{r}) \Big) + e^{i \sigma_0} \;  \times \\
        & \times
    \begin{cases}
        \frac{F_0 (\tilde{\eta}_0, \; \tilde{k}_0 r)}{F_0 (\tilde{\eta}_0, \; \tilde{k}_0 d)} \; \left[ \frac{F_0 (\eta, k d)}{k r} + f_0 (k) \; \frac{G_0 (\eta, k d) + i \: F_0 (\eta, k d)}{r} \right] - \frac{F_0 (\eta, \rho)}{k r}, \;\;\;\;\;\;\;\;\;\;\;\: r < d_0 \\
        f_0 (k) \; \frac{G_0 (\eta, \rho) + i \: F_0 (\eta, \rho)}{r}, \;\;\;\;\;\;\;\;\;\;\;\;\;\;\;\;\;\;\;\;\;\;\;\;\;\;\;\;\;\;\;\;\;\;\;\;\;\;\;\;\;\;\;\;\;\;\;\;\;\;\;\;\;\;\;\;\;\;\;\;\;\;\;\;\;\; r \geq d_0 .
    \end{cases}
    \end{split}
    \label{psi_pp_l0}
\end{equation}

This case is of a particular interest as the asymptotic part of the WF is equivalent to the so-called Lednicky-Lyuboshits (LL) formula \cite{Lednicky:2005tb, Lednicky:1981su} (i.e. $\psi_\mathrm{LL} (k, \; r) \equiv \left. \psi_{c+s}^{l=0} (k, \; r) \right|_{r>d_0} $), a model that has been commonly used in heavy-ion collision experiments in application to femtoscopy \cite{ALICE:2015hvw, Gos:2007cj}. 
We argue that using just the asymptotic part of the WF to calculate the femtoscopic CF is inappropriate\footnote{It's worth mentioning that similar to Eq. (\ref{psi_pp_l0}) form of the WF was discussed in \cite{Gmitro} but we couldn't find any experimental applications of it.} due to its singular behavior at small $r$: $\psi_\mathrm{LL} (k, \; r) \underset{ r \to 0}{\to} r^{-1}$.
Since the integration in Eq. (\ref{eq_CF_th_weightened_l01}) is performed in the whole coordinate space (including the small $r$ region) the usage of the LL model would cause an overestimation of the resulting CF.
On the other hand each partial wave term in the general solution obtained in this work (\ref{eq_Coul_str_sol_finite_waves}) behaves as $u_l (k, \; r) \underset{ r \to 0}{\to} r^{l+1} + o(r^{l+1})$ ensuring regularity of the WF at small relative distances $r$.

\begin{figure}[h]
    \centering
    \includegraphics[width=0.48\linewidth]{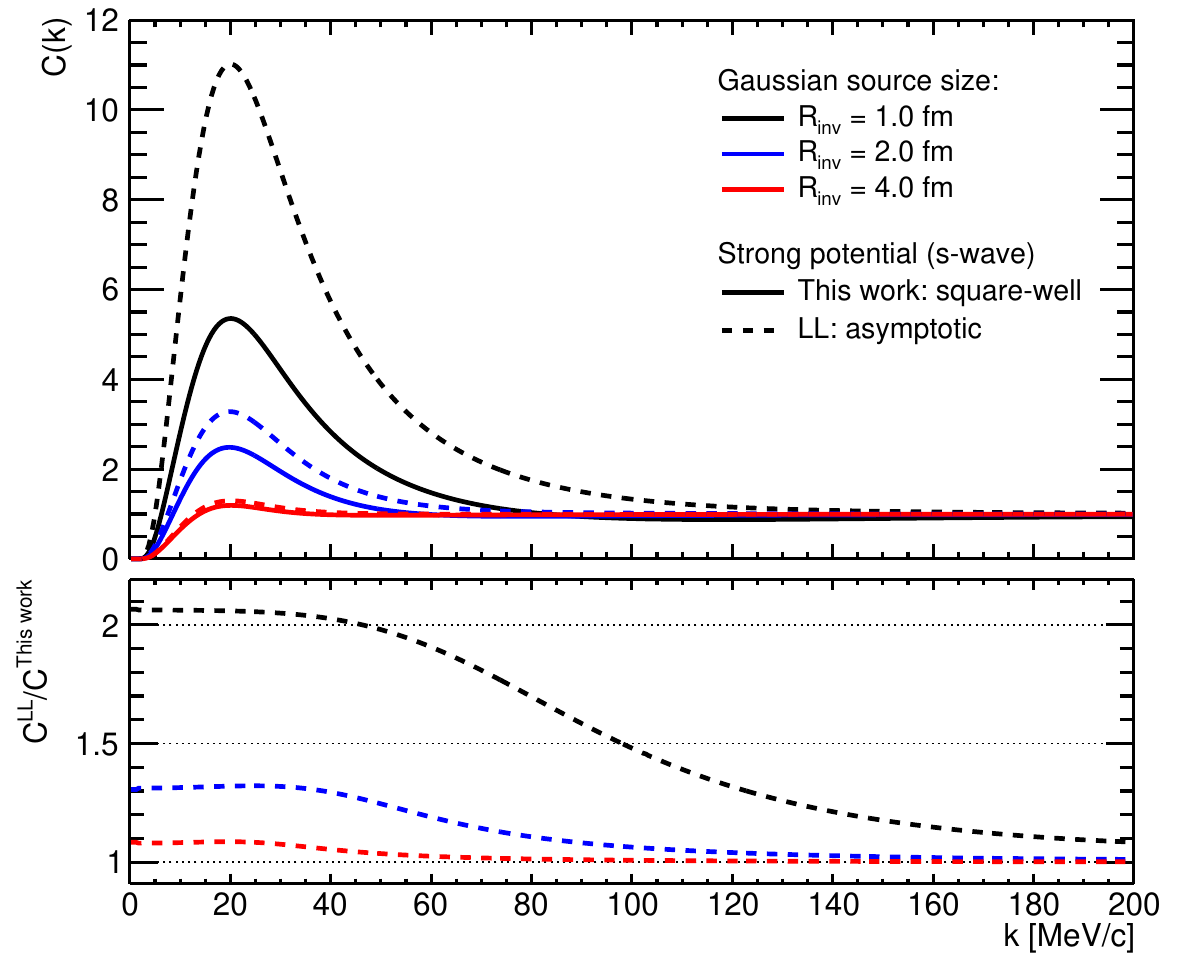}
    \includegraphics[width=0.48\linewidth]{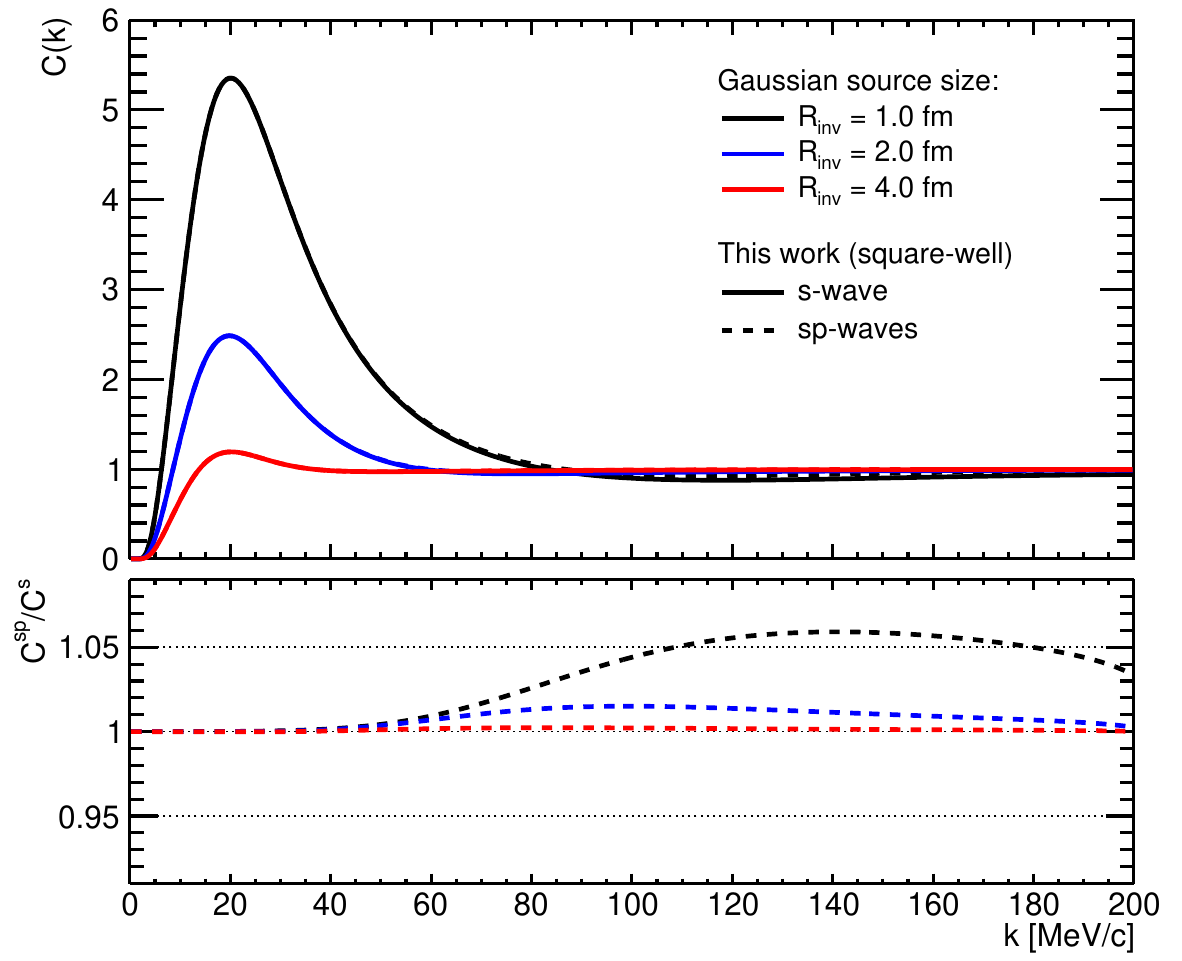}
    \caption{Proton-proton CFs calculated for different effective source sizes with the Coulomb+SW model. Left: the Coulomb+SW model accounting for strong interaction in the s-wave (continuous lines) is compared to the LL approach (dashed lines). Right: the s-wave calculation (continuous lines) with the Coulomb+SW model is compared to the sp-wave configuration (dashed lines).
    }
    \label{fig:comp1}
\end{figure}

A direct comparison of the proton-proton femtoscopic CFs calculated with the LL model and the one presented in this work is shown in Fig. \ref{fig:comp1} (left). As the shape of the correlation function depends on the source function $S(\Vec{r}, R_\mathrm{eff})$ (see Eq. (\ref{eq_CF_th}) and (\ref{eq_CF_th_weightened_l01})), the CFs are obtained for three different effective source sizes assuming a Gaussian profile. 
The choice of the values of $R_\mathrm{eff}$ is motivated by experimental results from proton pair femtoscopy at the LHC for different colliding systems, from proton-proton (pp) collisions with $R_\mathrm{eff} \approx 1$ fm \cite{ALICE:2018ysd,ALICE:2020ibs} to Pb--Pb collisions with $R_\mathrm{eff} \approx 4$ fm \cite{ALICE:2015hvw, ALICE:2025wuy}. We include an example for $R_\mathrm{eff} = 2$ fm to cover intermediate size systems. 
The comparison demonstrates that the LL model leads to a significant overestimation of the CF, as expected, which is more pronounced for smaller sources.

Moreover, compared to the LL model that is limited to the s-wave, our approach offers the possibility to include the strong potential in higher partial waves. This allows us to verify the dominance of the s-wave contribution by comparing directly the CFs that include only the s-wave against those that contain both s- and p-waves. Such a comparison is reported in the right panel of Fig. \ref{fig:comp1}, manifesting a discrepancy of only up to 6\% for $R_\mathrm{eff}=1$ fm in the $100 < k < 180$ MeV/c region. This is consistent with the aforementioned consideration that the strong potential can be safely neglected in higher order waves due to the presence of the centrifugal term in the Schr\"odinger's equation (\ref{eq_Schred_rad_1}).

\begin{figure}[h]
    \centering
    \includegraphics[width=0.48\linewidth]{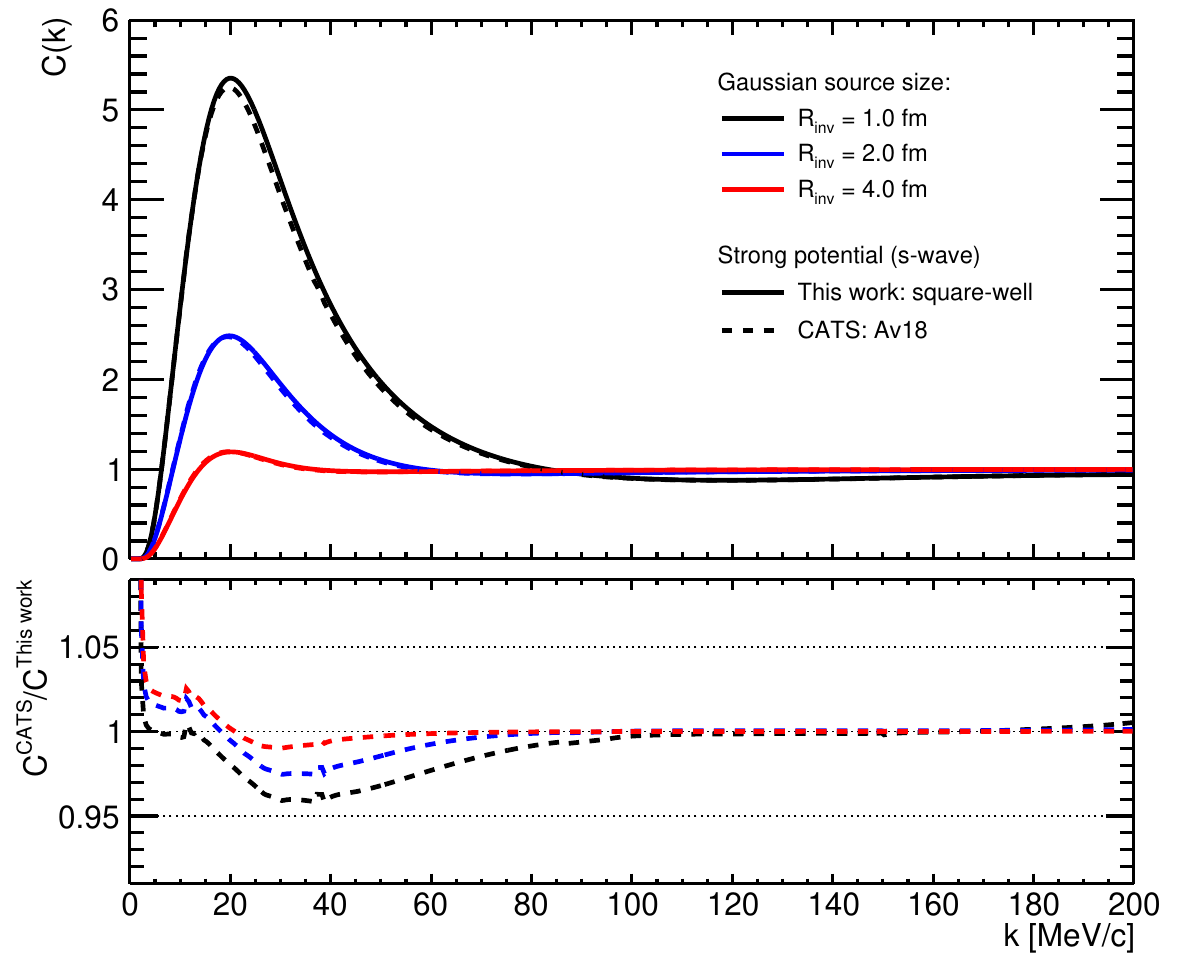}
    \includegraphics[width=0.48\linewidth]{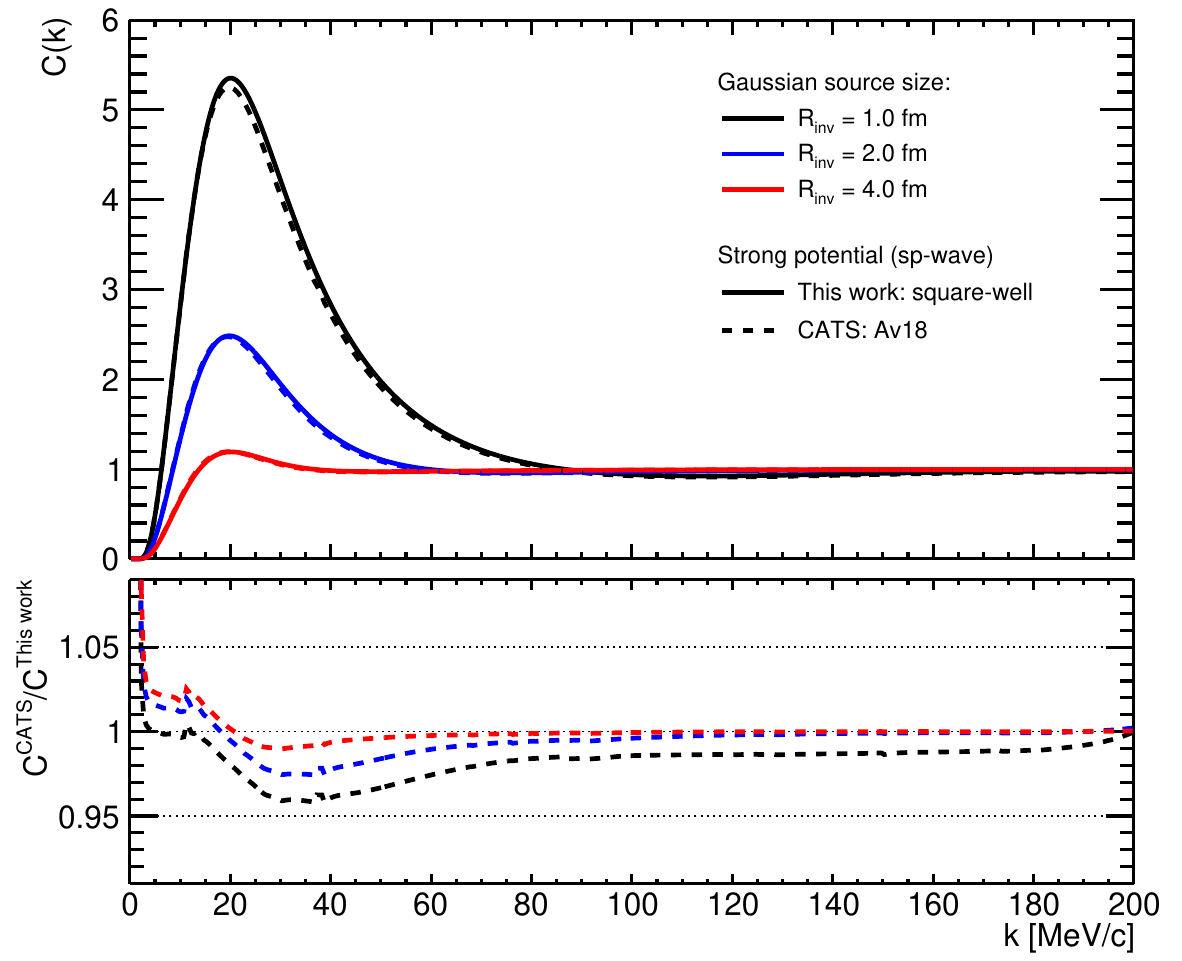}
    \caption{Proton-proton CFs with the CATS package employing the Argonne $v18$ strong potential model compared to the Coulomb+SW model for the two separate cases in which the strong potential is included in the s- (left) and sp-waves (right).}
    \label{fig:comp2}
\end{figure}

The inability of the LL model to be applied to small systems was observed also in Ref. \cite{ALICE:2018ysd} employing the CATS package \cite{Mihaylov:2018rva} that numerically solves the Schr\"odinger's equation for a given potential shape. 
Therefore, we compare our calculation with the output of CATS, used as a benchmark, profiting from the fact that it can operate with realistic strong potential models such as Nijm93, Reid93~\cite{PhysRevC.49.2950} and Argonne $v18$~\cite{PhysRevC.51.38} among others, and uses the partial wave extension. 
We obtained the proton-proton CFs with the CATS package employing the Argonne $v18$ strong potential model as it was chosen for recent applications \cite{ALICE:2018ysd, Mahlein:2023fmx, ALICE:2025wuy, DiMauro:2024kml} and it disproved the LL approach \cite{ALICE:2018ysd}. The models are compared in Fig. \ref{fig:comp2} for the two separate cases in which the strong potential is included in the s- (left) and sp-waves (right). The resulting CFs are rather close with a mild discrepancy in the peak region that is more noticeable for smaller source sizes. This indicates that the femtoscopic signal measured in small collision systems is prone to be more sensitive to the strong potential shape. Nevertheless, such a small difference might be too insignificant (up to $\approx 4 \%$ for the \mbox{$R_{inv} =1$ fm}) for the case of application to experimental data. 
For example, the total uncertainties (quadratic sum of statistical and systematic uncertainties) of the proton pair CF reported in Ref. \cite{ALICE:2018ysd} range from $\approx 6.7 \%$ at $k^* = 22$ MeV/c to $\approx 3.1 \%$ at $k^* = 98$ MeV/c exceeding the discrepancy between the CATS and the effective analytical calculations of this work. Moreover, the experimental distributions are always given in a discrete (binned) form, meaning that each experimental value of the CF corresponds to some range of $k$. This additional built-in uncertainty of the argument of the CF results in a higher tolerance to the discrepancy.

The agreement with the CATS calculations allows us to draw the following conclusions: (i) the proposed analytical effective treatment of the strong potential provides a reasonable and reliable description of the proton-proton femtoscopic CF; (ii) the agreement exposes that the femtoscopic observables might not be sensitive to the actual shape of the strong potential.
The QM problem for which we obtained the solution (\ref{eq_Coul_str_sol_all_waves}-\ref{eq_Coul_str_sol_finite_waves}) is a particle-on-particle scattering problem: no particular information about the shape of the strong potential (generally, of any short-range one) can be extracted from scattering data at low momenta/energy \cite{Bethe49}, as pointed out in the past. A similar argument might apply to the femtoscopic observables as the signal is present in the low relative momentum region.

Nonetheless, we believe that femtoscopy in heavy-ion collisions can still be a unique tool to study the strong interaction as one one could go beyond pairs and study three-body correlations \cite{ALICE:2022boj} or explore correlations between rare baryons (like $\Lambda$ and $\Xi$ \cite{ALICE:2020mfd}) that can be produced in sufficient amounts in high-energy hadronic collision experiments. Considering the latter, another unique advantage of our Coulomb+SW model arises. Being it fully analytical, the WF is obtained in a general form regardless of the square-well potential parameters (see Eq. \ref{eq_Coul_str_sol_finite_waves}).
This means that the measured femtoscopic CFs of pairs of (rare) hadrons could be fitted with the model to solve the ''reverse'' problem of defining the effective strong potential parameters from such a fit\footnote{In this work the effective range parametrization (\ref{eq_A2_7}) for phase shifts is used to account for the strong potential in the asymptotic part of the WF (\ref{eq_Coul_str_asympt_Led}), but in principle the definition of the whole WF can be reduced to only two parameters -- width $d_l$ and depth $V_l$ by using Eq. (\ref{eq_A2_5}).}. These parameters can provide an indirect effective estimate of the phase shifts at small energy scale since -- as mentioned before -- even a simple square-well can provide a decent description at low momenta.

\section{Conclusion}\label{sec13}

In this work we developed an analytical treatment of the short-range strong interaction in two-nucleon systems and explored its applicability to correlation femtoscopy in high-energy hadronic and nuclear collisions. 
By modeling the strong potential with a square-well form, we obtained a solution of the Schrödinger equation in the presence of the Coulomb interaction, yielding a pair wave function that is regular at small relative distances and can incorporate contributions from multiple partial waves. 
We applied the formalism to proton–proton femtoscopy and computed correlation functions for realistic Gaussian source sizes. A direct comparison with the Lednicky–Lyuboshits model showed that the usage of the latter leads to a noticeable overestimation of the correlation functions, thus questioning its applicability to the case of proton femtoscopy.
The comparison with numerical solutions obtained using the CATS framework and the Argonne $v18$ potential yielded a satisfactory agreement for all the considered effective sizes of a proton-emitting source.
These results also exposed that femtoscopic observables at low relative momentum may be insensitive to the detailed shape of the short-range strong potential.

Owing to its fully analytical nature, the approach presented here offers a practical and flexible tool for femtoscopic analyses of nucleon pairs. Future developments include the potential applicability of the model to systems where numerical solutions are computationally demanding or to those configurations for which the reverse problem needs to be solved, such as correlations involving rare baryons. 

\backmatter


\bmhead{Acknowledgements}
We thank Paolo Finelli and Francesco Noferini for useful discussions, as well as Dimitar Mihaylov for the assistance with the CATS framework.

\section*{Declarations}

This work has received funding from the European Research Council (ERC) under the European Union’s Horizon 2020 research and innovation programme, grant agreement No. 950692.



\begin{appendices}

\section{Solution for the Coulomb potential.}\label{secA1}

It is convenient to look for the solution $u_l (r)$ of the Eq. (\ref{eq_Schred_rad_1}) with $V(r) = \frac{\hbar^2}{\mu a_B r}$ in the form

\begin{equation}
    u_l (r) = \rho^{l+1} e^{i \rho} v_l (\rho) .
    \label{eq_solution_subst}
\end{equation}

Using the $z = - 2 i \rho$ substitution, the equation (\ref{eq_Schred_rad_1}) can be brought to a Kimmer-like one:

\begin{equation}
    z \; \frac{\mathrm{d}^2 v_l}{\mathrm{d} z^2} + \Big( 2 (l + 1) - z \Big) \; v_l - (l + 1 + i \eta) \; v_l = 0 .
    \label{eq_Kimmer}
\end{equation}

In Mathematics the Kimmer's equation has \textit{regular} and \textit{singular} solutions \cite{Abramowitz_Stegun:1972}:

\begin{enumerate}
    \item Kummer's (confluent hypergeometric) function $_1F_1(a, b, z)$ ,
    \item Tricomi's (confluent hypergeometric) function $U(a, b, z)$.
\end{enumerate}

Choosing the first option and applying it to Eq. (\ref{eq_solution_subst}) one obtains the regular (at $r=0$) solution to Eq. (\ref{eq_Schred_rad_1}) (with the Coulomb potential) called \textit{regular Coulomb wave function}:

\begin{equation}
    F_l (\eta, \rho) := u_l^{reg}  = C_l \rho^{l+1} e^{i \rho} _1F_1( l + 1 + i \eta, \; 2l+2, \; - 2 i \rho),
    \label{eq_Coul_reg}
\end{equation}

\noindent where $C_l = \frac{2^l e^{- \frac{\pi \eta}{2}} | \Gamma (l + 1 + i \eta) |}{(2l + 1)!}$ is the \textit{Coulomb normalization constant} needed to satisfy required asymptotics $F_l \underset{r \to \infty}{\sim} \sin \: (\rho - \eta \ln (2 \rho) - l \frac{\pi}{2} + \sigma_l)$. \newline


Irregular solutions to Eq. (\ref{eq_Schred_rad_1}) are given in terms of \textit{irregular Coulomb wave functions} $u_l^{(\pm)} (\eta, \rho)$\footnote{Irregular Coulomb wave functions $u_l^{(\pm)}$ are expressed via the Tricomi's (confluent hypergeometric) function $U(a, b, z)$ \cite{Messiah}.} and $G_l (\eta, \rho)$ that have the following asymptotics \cite{Messiah}

\begin{equation}
    \begin{split}
        & u_l^{(\pm)} \underset{r \to \infty}{\sim} e^{\pm i (\rho - \eta \; \ln \: 2 \rho - \frac{l \pi}{2} ) } , \\
        & G_l \underset{r \to \infty}{\sim} \sin \: (\rho - \eta \ln (2 \rho) - l \frac{\pi}{2} + \sigma_l) .
    \end{split}
    \label{eq13}
\end{equation}

Relations between the regular and irregular Coulomb wave functions are given as \cite{Messiah}

\begin{equation}
    \begin{split}
        & u_l^{(\pm)} = e^{\mp i \sigma_l} \; \Big( G_l \pm i \: F_l \Big) , \\
        & u_l^{(-)} = {u_l^{(-)}}^* , \\
        & F_l = \frac{1}{2i} \; \Big( u_l^{(+)} e^{i \sigma_l} - u_l^{(-)} e^{- i \sigma_l} \Big) , \\
        & G_l = \frac{1}{2} \; \Big( u_l^{(+)} e^{i \sigma_l} + u_l^{(-)} e^{- i \sigma_l} \Big) .
    \end{split}
    \label{eq14}
\end{equation}

\section{Matching the solutions.} \label{secA2}

The total sought-after solution (\ref{eq_sol_gen}) for the case of Coulomb plus square-well potential (Fig. \ref{fig1}) is conveniently rewritten in the following form

\begin{equation}
    \psi (\Vec{r}) = \frac{1}{r} \sum_{l=0}^\infty i^l (2l + 1) u_l' (r) P_l (\cos \theta),
    \label{eq_A2_1}
\end{equation}

\noindent with

\begin{equation}
    u_l' (r) =
    \begin{cases}
        A_l \; {u_l'}^{(I)} (r), \;\;\;\;\;\; r < d_l , \\
        {u_l'}^{(II)} (r), \;\;\;\;\;\;\;\;\;\; r \geq d_l ,
    \end{cases}
    \label{eq_A2_2}
\end{equation}

\noindent where $A_l$ is a normalization constant that must be obtained from the matching condition at $r=d_l$.

Using the solutions (\ref{eq_Coul_str_sol_reg1}) and (\ref{eq_Coul_str_asympt_Led}) for the regions (I) and (II) respectively (see Fig. \ref{fig1}) one defines ${u_l'}^{(I)}$ and ${u_l'}^{(II)}$:

\begin{equation}
    \begin{split}
        & {u_l'}^{(I)} (r) = \frac{1}{\tilde{k}_l} \; e^{i \tilde{\sigma}_l} F_l (\tilde{\eta}_l, \; \tilde{\rho}_l) , \\
        & {u_l'}^{(II)} (r) = e^{i \sigma_l} \Big( \frac{F_l (\eta, \rho)}{k} + f_l (k) \; \big(G_l (\eta, \rho) + i \: F_l (\eta, \rho) \big) \Big) .
    \end{split}
    \label{eq_A2_2}
\end{equation}

The final solution must be continuous and smooth (differentiable) at $r=d_l$ for each $l$ which implies the following matching conditions:

\begin{equation}
    \begin{split}
        & A_l \; {u_l'}^{(I)} (r = d_l) = {u_l'}^{(II)} (r = d_l) , \\
        & {\left. \frac{\mathrm{d}}{\mathrm{d} r} \Big( \ln \: ( A_l \; {u_l'}^{(I)} (r) ) \Big) \right|}_{r=d_l} = {\left. \frac{\mathrm{d}}{\mathrm{d} r} \Big( \ln \: ( {u_l'}^{(II)} (r) ) \Big) \right|}_{r=d_l} .
    \end{split}
    \label{eq_A2_3}
\end{equation}

The first one is used to define the normalization constant $A_l$:

\begin{equation}
    A_l = \frac{\tilde{k}_l}{F_l (\tilde{\eta}_l, \; \tilde{k}_l d)} \; e^{i (\sigma_l - \tilde{\sigma}_l)} \; \Big( \frac{F_l (\eta, k d)}{k} + f_l (k) \; \big(G_l (\eta, k d) + i \: F_l (\eta, k d) \big) \Big) .
    \label{eq_A2_4}
\end{equation}

The second one describes the dependence of the phase shifts associated with the short-range potential on the pair relative momentum $k$:

\begin{equation}
    \cot \delta_l = \frac{G_l (\eta, k d)}{F_l (\eta, k d)} \; \frac{\tilde{k} \; f_l ( \tilde{\eta}, \; \tilde{k} d ) - k \; g_l (\eta, k d)}{k \; f_l (\eta, k d) - \tilde{k} \; f_l ( \tilde{\eta}, \; \tilde{k} d )} ,
    \label{eq_A2_5}
\end{equation}

\noindent where the $f_l$ and $g_l$ are the logarithmic derivatives of $F_l$ and $G_l$ respectively:

\begin{equation}
    \begin{split}
        & f_l (\eta, \; \rho) = \frac{\mathrm{d}}{\mathrm{d} r} \Big( \ln \: ( F_l (\eta, \; \rho) ) \Big) , \\
        & g_l (\eta, \; \rho) = \frac{\mathrm{d}}{\mathrm{d} r} \Big( \ln \: ( G_l (\eta, \; \rho) ) \Big) .
    \end{split}
    \label{eq_A2_6}
\end{equation}



\section{Potential and effective range parameters.} \label{secA3}

\begin{figure}[h!]
    \centering
    \includegraphics[width=0.45\linewidth]{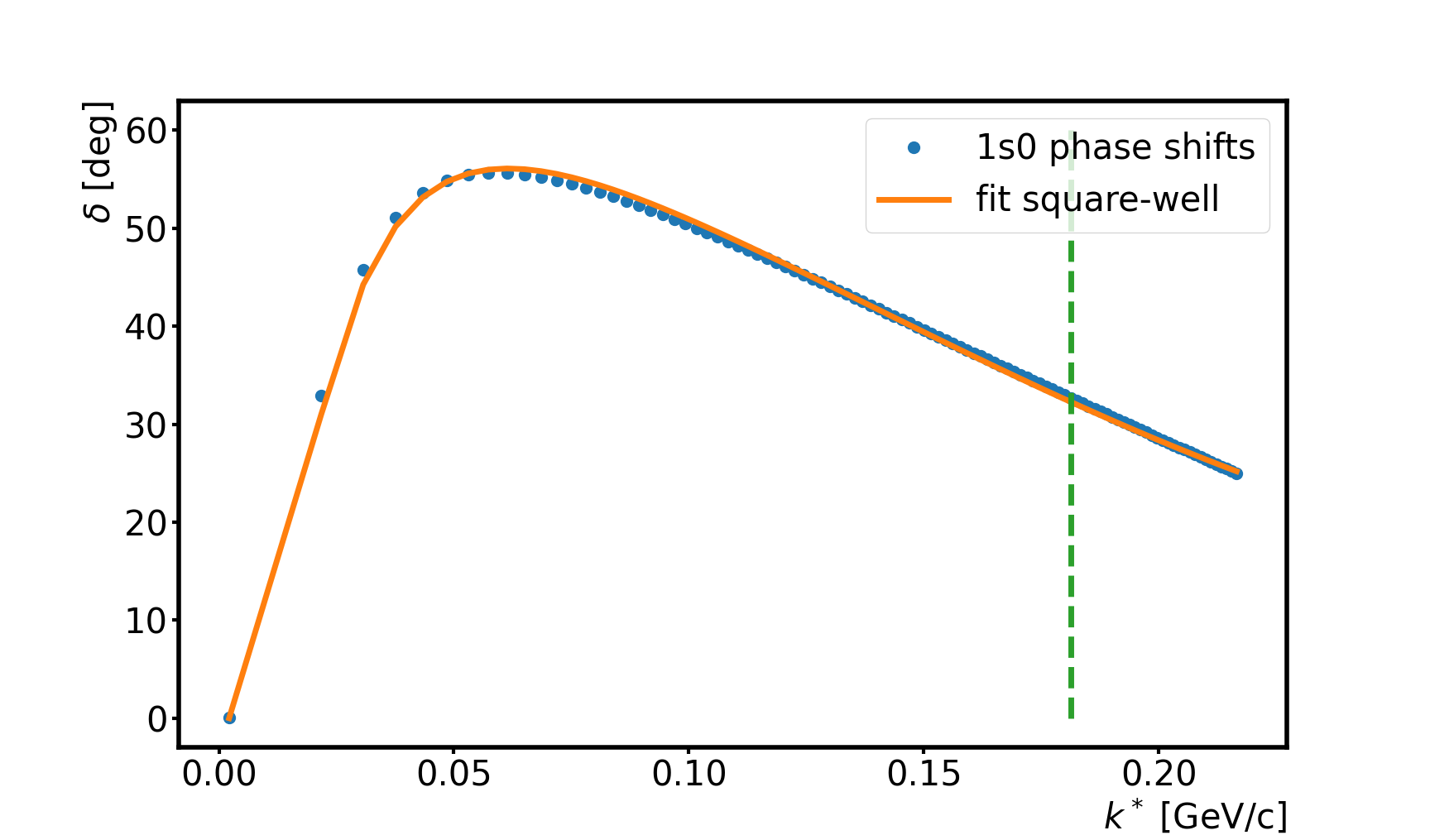}
    \includegraphics[width=0.45\linewidth]{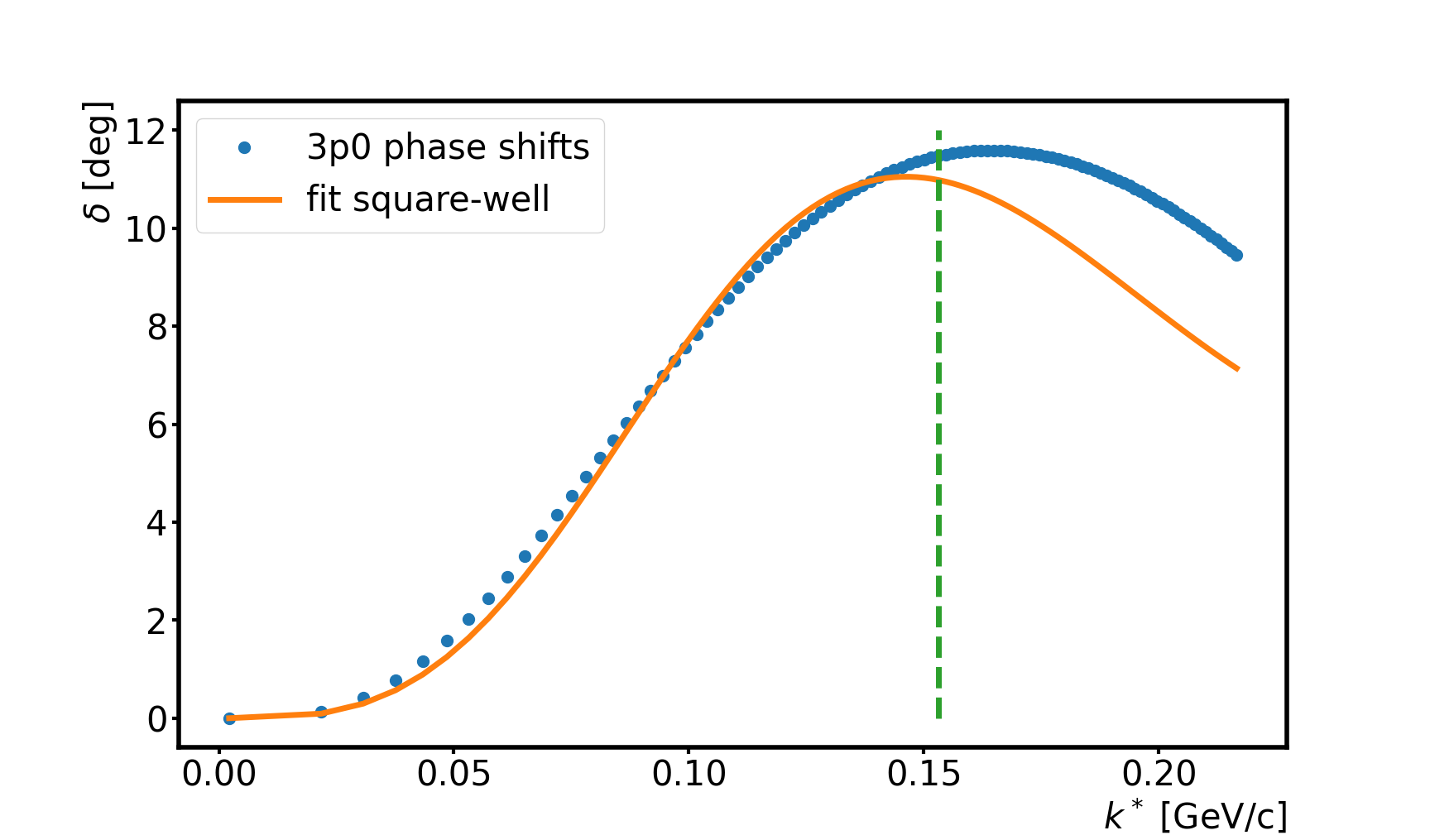}
    \includegraphics[width=0.45\linewidth]{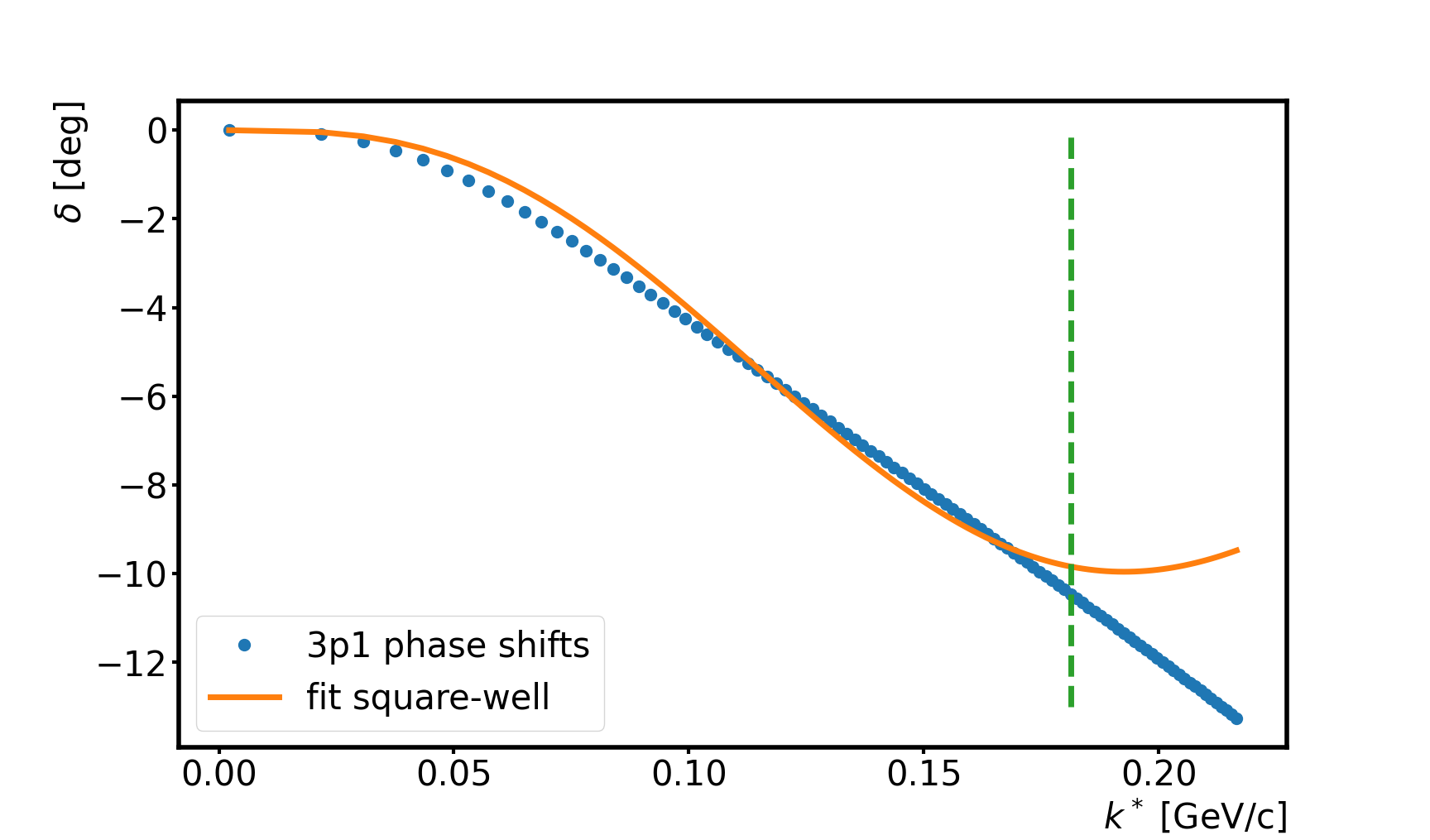}
    \includegraphics[width=0.45\linewidth]{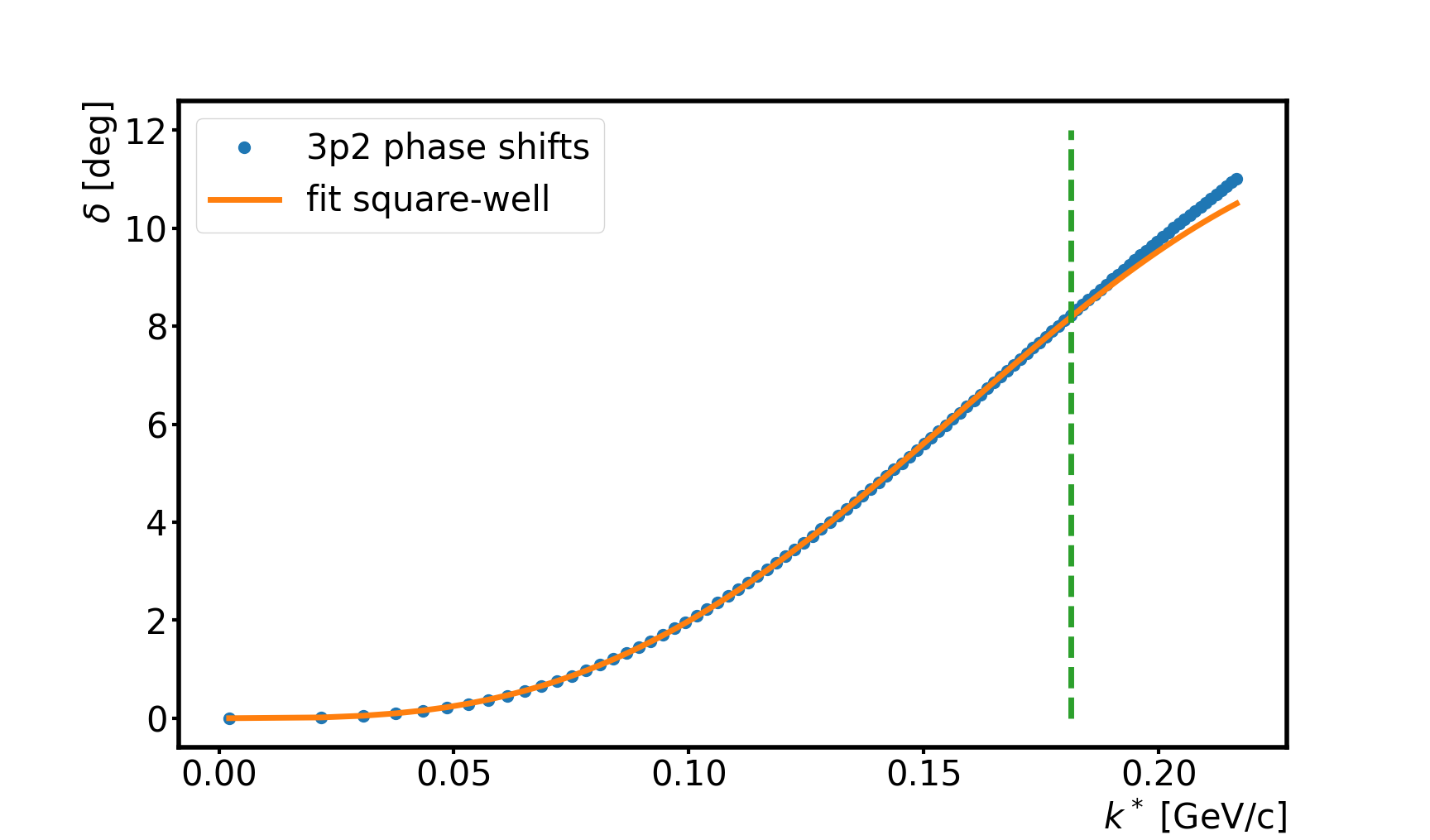}
    \caption{p-p phase shifts fitted with Eq. (\ref{eq_A2_5}) to extract the effective strong potential parameters (Tab. \ref{tab1}) in a shape of a square-well for different pair states: $^1S_0$ (top left), $^3P_0$ (top right), $^3P_1$ (bottom left), $^3P_2$ (bottom right). The green line corresponds to the upper limit of the fit range.}
    \label{fig_A3_1}
\end{figure}

The square-well parameters for the strong potential are extracted via a fit of the phase shift data from proton-proton scattering experiments \cite{Stoks:1993tb}\footnote{In this work we used the Nijmegen database (https://nn-online.org) to export the phase shift data based on the partial-wave analysis of Ref. \cite{Stoks:1993tb}} with Eq. (\ref{eq_A2_5}). In this work we confine the number of considered pair states to the first two values of pair orbital momentum $L$ including the spin-degeneracy for $L=1$. The extracted parameters are listed in Tab. \ref{tab1} and the fits are presented in Fig. \ref{fig_A3_1}.

\begin{table}[h]
\caption{Effective parameters of the strong potential in a shape of a square-well extracted from the fits of experimental phase shifts (Fig. \ref{fig_A3_1}) with Eq. \ref{eq_A2_5}.}\label{tab1}%
\begin{tabular}{@{}llll@{}}
\toprule
L & Pair's state $^{2S+1}L_J$  & Depth $\frac{2 \mu}{\hslash^2} V_{^{2S+1}L_J}$ [GeV] & Width $d_{^{2S+1}L_J}$ [fm]\\
\midrule
0 & $^1S_0$ & $-0.01620$ & $2.30394$ \\
1 & $^3P_0$ & $-0.00283$ & $4.03726$ \\
1 & $^3P_1$ & $0.00374$ & $3.43222$ \\
1 & $^3P_2$ & $-0.01023$ & $2.18680$ \\
\botrule
\end{tabular}
\end{table}

To take into account the influence of the strong potential in the asymptotic region (II) we used the scattering amplitude in the form given by Eq. (\ref{eq_Scat_ampl}). The effective range parametrization for the cotangent term for the simplest case of $l=0$ and absence of Coulomb interaction is well known \cite{Bethe49}. However, as it does not apply to out case we utilize the solution given in Ref. \cite{VANHAERINGEN1981317}, which takes into account Coulomb interaction and is defined for partial waves with $l>0$. It is given as

\begin{equation}
    k^{2l+1} \Big[ A_c(\eta) \big( \cot \delta_l -i \big) + 2 \eta H(\eta) \Big] \prod_{n=1}^l \left( 1 + \frac{\eta^2}{n^2} \right) = - \frac{1}{f_l} + \frac{1}{2}d_l k^2 + (-P d_l^3 k^4) + O(k^6) ,
    \label{eq_A2_7}
\end{equation}

\noindent where

$$H(\eta) \equiv \psi (i \eta) + \frac{1}{2 i \eta} - \ln \left( -i \eta \: \sign \left( - \frac{1}{a_B} \right) \right) ,$$

\noindent and $\psi(x)$ here is the digamma function. \newline

\begin{figure}
    \centering
    \includegraphics[width=0.45\linewidth]{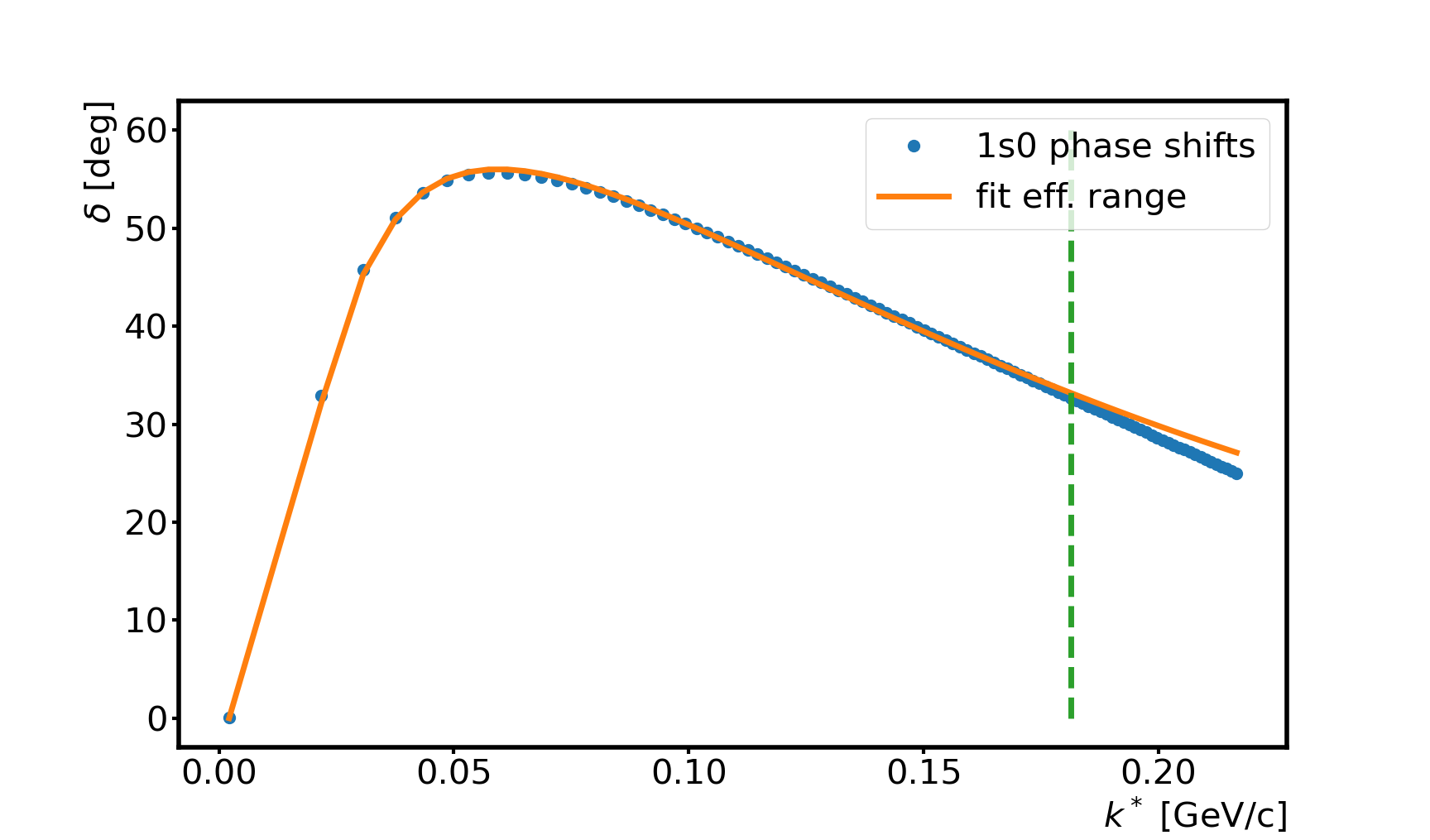}
    \includegraphics[width=0.45\linewidth]{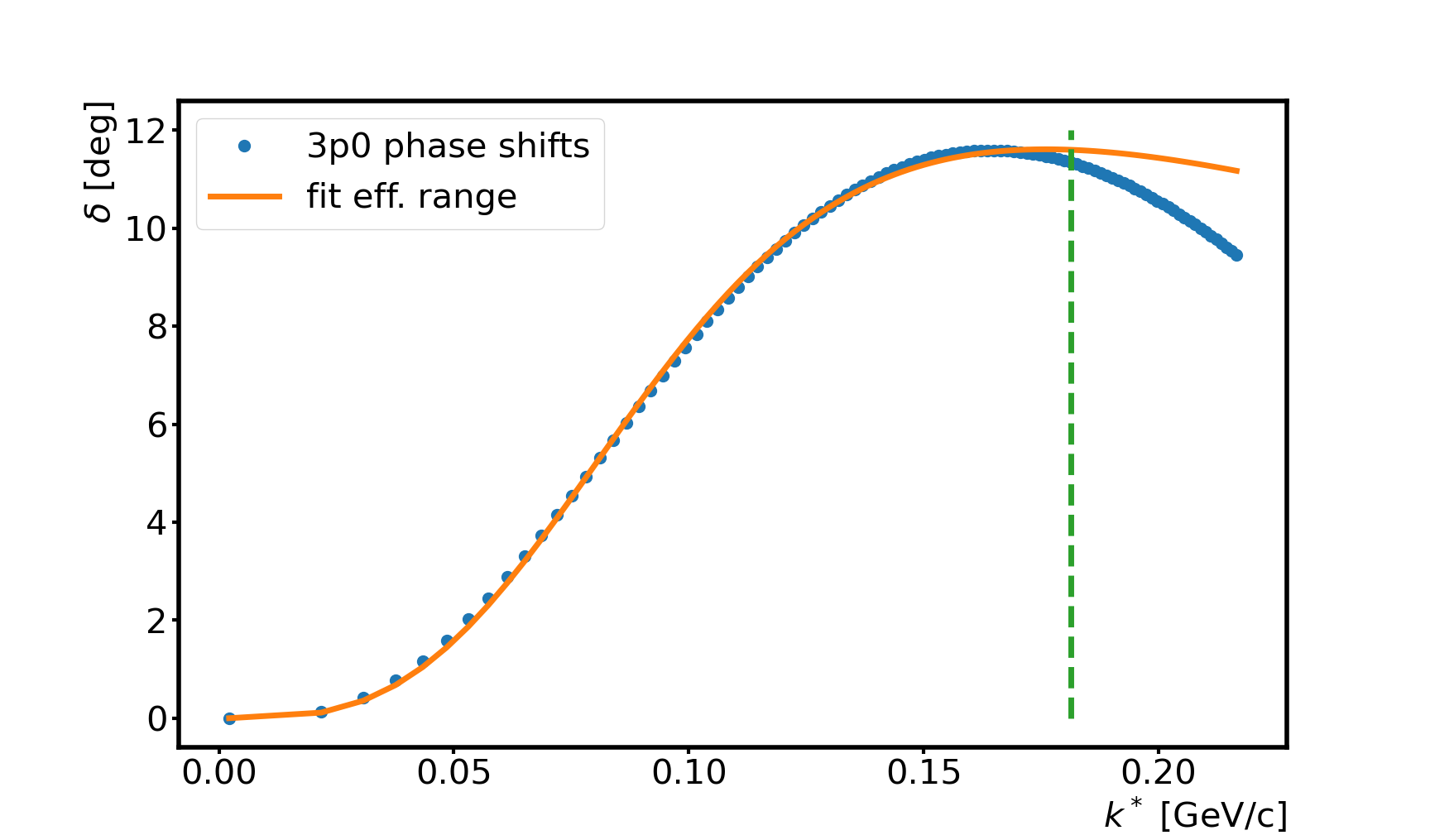}
    \includegraphics[width=0.45\linewidth]{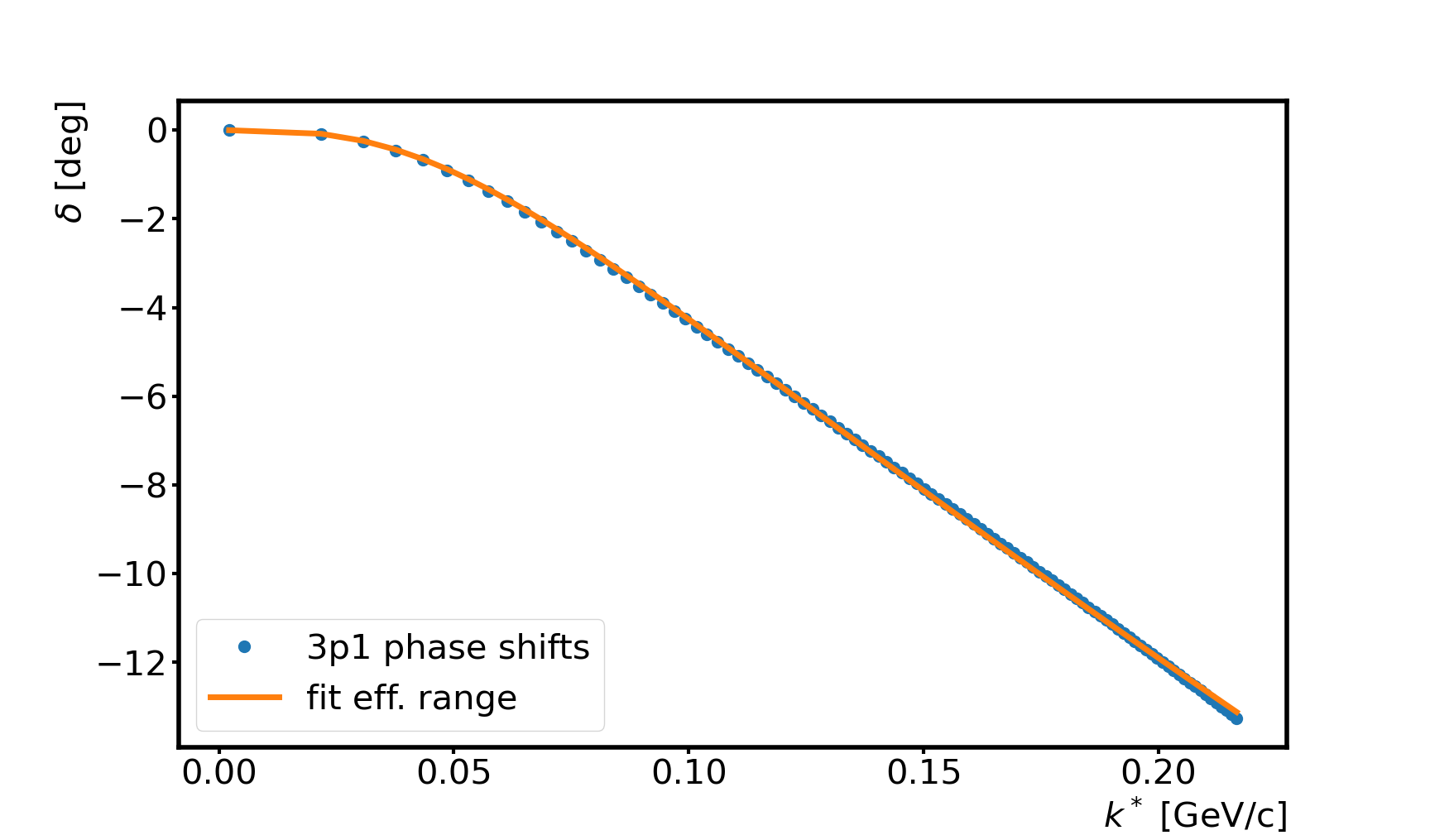}
    \includegraphics[width=0.45\linewidth]{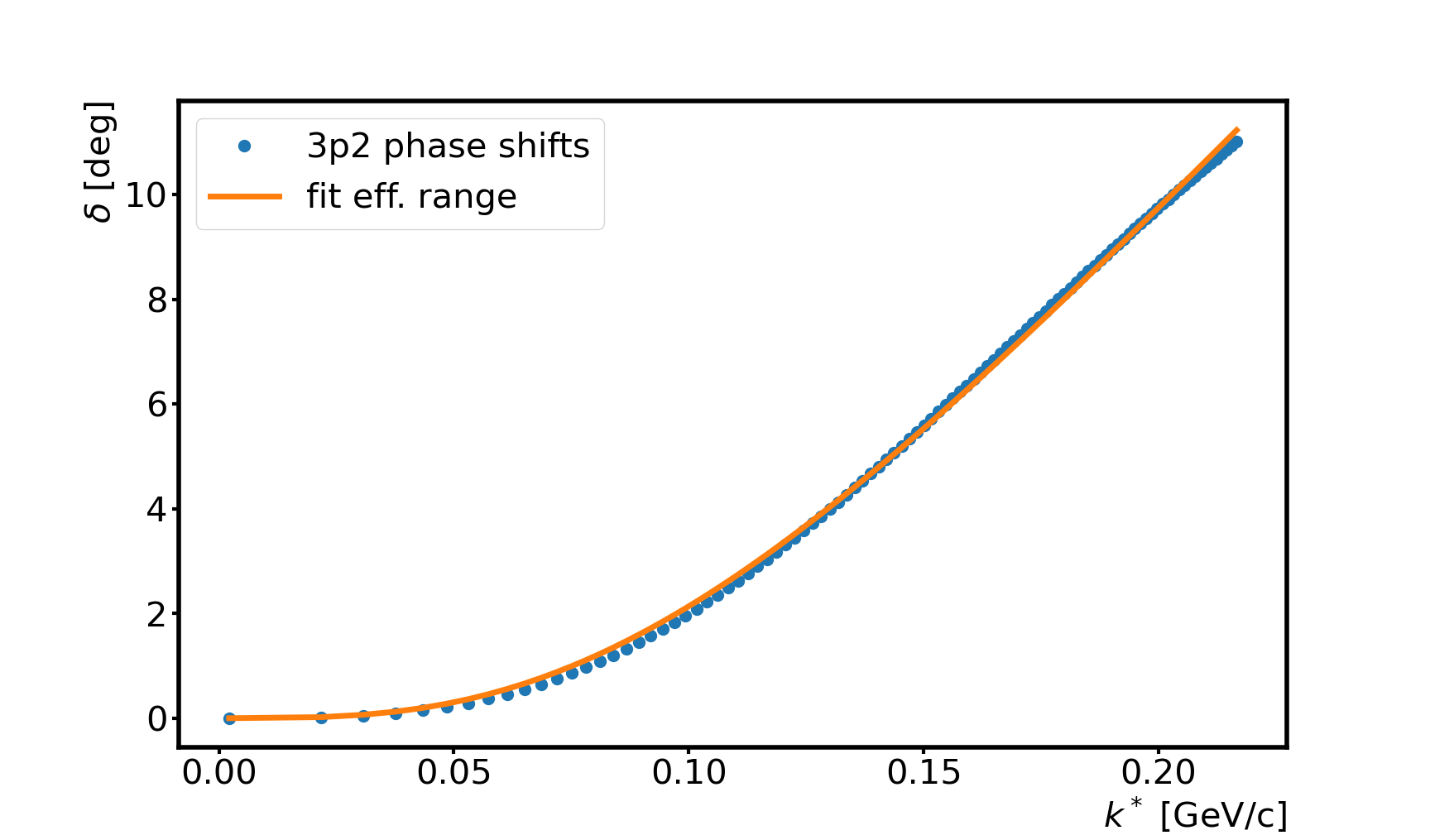}
    \caption{p-p phase shifts fitted with Eq. (\ref{eq_A2_7}) to extract the effective range parameters (Tab. \ref{tab2}) for different pair states: $^1S_0$ (top left), $^3P_0$ (top right), $^3P_1$ (bottom left), $^3P_2$ (bottom right). Green line corresponds to the upper limit of the fit range.}
    \label{fig_A3_2}
\end{figure}

This equation us used to parametrize the $\cot \delta_l$ needed to account for the strong interaction in the asymptotic region (see Eq. (\ref{eq_Coul_str_asympt_Led}-\ref{eq_Scat_ampl})). The extracted parameters are listed in Tab. \ref{tab2} and the fits are presented in Fig. \ref{fig_A3_2}.

\begin{table}[h]
\caption{Effective range parameters $f_l$, $d_l$ and $P_l$ used in the parametrization (\ref{eq_A2_7}) of the scattering amplitude to account for the strong potential in the asymptotic region.}\label{tab2}%
\begin{tabular}{@{}lllll@{}}
\toprule
L & Pair's state $^{2S+1}L_J$ & $f_{^{2S+1}L_J}$ [fm] & $d_{^{2S+1}L_J}$ [fm] & $P_{^{2S+1}L_J}$\\
\midrule
0 & $^1S_0$ & $-7.52584$ & $2.46092$ & $-0.02547$ \\
1 & $^3P_0$ & $-2.48927$ & $2.00572$ & $-0.41956$ \\
1 & $^3P_1$ & $1.86182$ & $-7.81453$ & $0$ (fixed) \\
1 & $^3P_2$ & $-0.44475$ & $7.01762$ & $0$ (fixed) \\
\botrule
\end{tabular}
\end{table}





\end{appendices}

\newpage
\bibliography{sn-bibliography}

\end{document}